\documentclass[aps,prd,onecolumn,nofootinbib,superscriptaddress]{revtex4}
\usepackage{float}
\usepackage{graphicx}%Include figure files
\usepackage{amsmath}
\usepackage{amsfonts}
\usepackage{amssymb,ulem}
\usepackage{color}%
\usepackage{dcolumn}
\usepackage{subfigure}
\usepackage{pdfpages}
\usepackage{multirow}
\usepackage{MnSymbol,wasysym}
\usepackage{braket,diagbox}
\usepackage{eurosym}
\usepackage{calrsfs}
\usepackage[usenames,dvipsnames,svgnames]{xcolor}

\newcommand{\RNum}[1]{\uppercase\expandafter{\romannumeral #1\relax}}
\usepackage[colorlinks=true,linkcolor=blue,urlcolor=blue,filecolor=black,citecolor=red,
pdfstartview=FitV,pdftitle={},pdfsubject={},pdfkeywords={},pdfpagemode=None,bookmarksopen=true]{hyperref}

\begin{document}
\baselineskip=0.5 cm

\title{Timelike bound orbits and pericenter precession around black hole with conformally coupled scalar hair}

\author{Qi Qi}
\email{qiqiphy@163.com}
\affiliation{Center for Gravitation and Cosmology, College of Physical Science and Technology, Yangzhou University, Yangzhou, 225009, China}

\author{Xiao-Mei Kuang}
\email{xmeikuang@yzu.edu.cn (corresponding author)}
\affiliation{Center for Gravitation and Cosmology, College of Physical Science and Technology, Yangzhou University, Yangzhou, 225009, China}

\author{Yong-Zhuang Li}
\email{liyongzhuang@just.edu.cn}
\affiliation{Research center for theoretical physics, School of Science, Jiangsu University of Science and Technology, Zhenjiang, China.}

\author{Yu Sang}
\email{sangyu@yzu.edu.cn}
\affiliation{Center for Gravitation and Cosmology, College of Physical Science and Technology, Yangzhou University, Yangzhou, 225009, China}

\date{\today}

\begin{abstract}
\baselineskip=0.5 cm
We investigate the geodesic motions of timelike particles around a static hairy black hole with conformally coupled scalar field.
We mainly focus on the effects of the scalar charge and electric charge on the marginally bound orbits (MBO), innermost stable circular orbits (ISCO) and on the precessing orbits around this black hole. Our results show that both the scalar and electric charges suppress the energy as well as the angular momentum of the particles in the bound orbits. Then, we study the relativistic periastron precessions of the particles and constrain the charge parameters by employing the observational result of the S2 star's precession in SgrA*. It is found that the constraints on the
charge parameters from S2 star's motion are tighter than those from black hole shadow. Finally, we analyze the periodic motions of the particles and figure out samples of periodic orbits' configurations around the hairy black hole.
\end{abstract}

%%%%%%%

\maketitle
\tableofcontents
\newpage

\section{Introduction}
Black hole provides natural platform to test gravity in the strong field regime. Recent breakthroughs in astrophysical observation including gravitational waves (GW) \cite{LIGOScientific:2016aoc,LIGOScientific:2018mvr,LIGOScientific:2020aai}, black hole shadows \cite{EventHorizonTelescope:2019dse,EventHorizonTelescope:2019ths,EventHorizonTelescope:2019pgp,
EventHorizonTelescope:2022wkp,EventHorizonTelescope:2022xqj}  and the supermassive object in the center of our Milky Way \cite{Genzel:2010zy} give strong evidences of the existence of black holes in our Universe.
Those observations further demonstrate the success of Einstein's general relativity (GR), but the uncertainties in the data also leave some space for alternative theories of gravity. Therefore,
plenty of research on the properties in the vicinity of black holes have been carried on since those remarkable progresses provide potential tools to test or distinguish GR and modified gravity theories. 
Among them, the structures of geodesics for massive particles or photons  around the black holes have been attracting  extensive interest, since the orbits of photons are the first step to study the black hole shadow related physics, while the bound timelike orbits  are key ingredients to study the motions of stars around SgrA* as well as GW radiation during the black hole merger. Those studies could also inversely help us further understand the essential structure of gravity from the central black holes. 
From the timelike geodesic, one important type of bound  orbit is the precessing orbit. It is what the Mercury undergoes, which gives one of the early evidences of GR \cite{Willbook}. Later, the precessing of the stars orbiting around the supermassive black hole in the central of SgrA* has been disclosed  in \cite{Iorio:2011zi,Grould:2017bsw,DeLaurentis:2018ahr,GRAVITY:2019tuf,DES:2019ltu}.  The measurement of the precessing orbit of S2 star \cite{GRAVITY:2020gka} has then been fitting in different theories of gravity  to constrain the model parameters, see for examples \cite{Hees:2017aal,DellaMonica:2021xcf,Yan:2022fkr,DellaMonica:2023dcw}. Thus, these precessing orbits may serve as  potential tools to test the alternative theories of gravity.
Another important type of bound orbit is the periodic orbit. It is known that the generic  orbits can be treated as small deviations from periodic orbits which encode fundamental information about orbits around a central black hole \cite{Levin:2008mq}.  Especially, a  zoom-whirl classification for the periodic orbits of the massive particles was proposed in  \cite{Levin:2008mq}, which is characterized by three topological integers $(z, w, v)$ representing zoom, whirl and vertex behaviors of the orbit, respectively.  It was previously found in \cite{Grossman:2011im} that the periodic orbits with zoom-whirl pattern may be helpful for faster computation of adiabatic extreme mass ratio inspiral process, which more recently was  further testified in two models  \cite{Tu:2023xab,Li:2024tld}. It is believed that periodic orbits could give more significant information about strong-field properties of the spacetime than the precessing orbits, so the taxonomy  of the periodic orbits for timelike particles has inspired considerable extensive studies \cite{Tu:2023xab,Levin:2008ci,Levin:2009sk,Healy:2009zm,Misra:2010pu,Pugliese:2013xfa,Babar:2017gsg,
Wei:2019zdf,Rana:2019bsn,Wang:2022tfo,Mummery:2022ana,
Habibina:2022ztd,Deng:2020yfm,Lin:2021noq,Lin:2023rmo,Bambhaniya:2020zno,
Zhou:2020zys,Zhang:2022zox,Azreg-Ainou:2020bfl,Li:2023djs,Wu:2023wld,Lim:2024mkb,Li:2024tld} and references therein.

On the other hand,  it is commonly accepted that  a more general theory of gravity beyond GR is eagerly required,  and so lots of modified gravitational theories have been proposed \cite{Clifton:2011jh,Bakopoulos:2020mmy,Corelli:2020hvr}.  One of the remarkable ways is to introduce an additional scalar field  into the action of GR. 
The underlying motivations mainly stem from three aspects: (i)  scalar field is a potential candidate for dark matter, dark energy and inflation which were proposed to explain some observations of our Universe, but their essences are still  unclear. (ii) scalar field may be a kind of ubiquitous composite in nature, for example, the ultralight axions are indispensable in string theory \cite{Svrcek:2006yi}.  (iii) the hypothesis of no hair theorem in classical GR states that black holes have only three characterized quantities, i.e., the black hole mass, electric charge and angular momentum \cite{Israel:1967wq,Hawking:1971vc}. However, the no hair theorem of black hole is still not be verified and from the current observation there is no hint for the absence of other fundamental quantity describing black holes. So  the introducing of scalar field into the action could be a direct way to testify it. A minimally coupled scalar field with gravity  usually does not obey the Gauss-law,  so in this case a black hole cannot
have a non-trivial regular scalar hair \cite{Janis:1968zz} such that the no hair theorem holds. However, the output is different when the non-minimal couplings between the gravity and  scalar field are introduced. As the first counterexample to the no hair theorem using scalar field, the authors of \cite{Zou:2019ays,Bekenstein:1974sf} found that an Einstein-conformally coupled scalar theory could lead to a secondary scalar hair around the Bocharova-Bronnikov-Melnikov-Bekenstein (BBMB) black hole which was numerically reproduced in \cite{Myung:2019adj}. But the scalar field in this sector diverges at the horizon which is difficult to  be physically acceptable. This disadvantage  was then erased by  introducing a cosmological constant into the solution, named as  the ``Martinez-Troncoso-Zanelli (MTZ)" black hole, in which the scalar field singularity was pushed  behind the event horizon \cite{Martinez:2005di,Martinez:2002ru}. It is noted that the MTZ black hole only has a spherical or hyperbolic horizon depending on the sign of the cosmological constant,  but the planar solution is not allowed, which inspired some extensive studies \cite{Bardoux:2012aw,Bardoux:2013swa,Cisterna:2018hzf}. 
In particular, it is natural to add the Maxwell field into Einstein-conformally coupled scalar theory and extend the theory into Einstein-Maxwell-conformally
coupled scalar theory of which the action is \cite{Martinez:2005di}
%%%%%%%
\begin{equation}\label{eq:SZYL}
S=\frac{1}{2\kappa}\int{d^4x}\sqrt{-g}[R-F_{\mu\nu}F^{\mu\nu}-\frac{\kappa}{6}(\varphi^2 R+6\partial_{\mu}\varphi\partial^{\mu}\varphi)],
\end{equation}
where $\kappa=8\pi G$, $R$ is the Ricci scalar, $F^{\mu\nu}$ is the strength of Maxwell field $A_{\mu}$ and $\varphi$ is the scalar field.  The author of \cite{Astorino:2013sfa} then constructed a static spherically symmetric black hole with the conformally coupled scalar field  which we will review in next section. Many physical phenomena on this black hole have been extensively studied, for instance,  the stability of the black hole against perturbations \cite{Chowdhury:2018pre,Zou:2019ays}, the Hawking radiation of charged particles \cite{Chowdhury:2019uwi}, the black hole shadow  \cite{Khodadi:2020jij}, the gravitational lensing effects \cite{QiQi:2023nex} and photon ring bound \cite{Myung:2024pob}, in which the conformally coupled scalar field reflects significant influences. 

Thus, the aim of this paper is to study the bound timelike geodesics, precessing and periodic orbits around the static spherically symmetric black hole with the conformally coupled scalar field to further understand the gravitational aspects of the corresponding theory of gravity. Starting for the timelike geodesic equation in this black hole, 
we will first analyze the effects of the scalar charge and electric charge on the characterized quantities of the marginally bound orbits and innermost stable circular orbits. We will see that the energy and the angular momentum of the particles in the bound orbits are both suppressed by the larger charge parameters of the hairy black hole. Then, we shall calculate the relativistic periastron precessions of the particles, and connect the obtained precession angle with S2 star's precession in SgrA* to constrain the charge parameters, which we will find is tighter than those from black hole shadows addressed in \cite{Khodadi:2020jij}. Finally, we will move on to the periodic motions of the particle and analyze periodic orbits' configurations around the black hole. Our current studies could be helpful for us to better understand the gravitational structure of black hole with scalar and electric hairs.

The remaining of this paper is organized as follows. In section \ref{sec:bound orbots}, starting from the timelike geodesic, we will analyze the properties of bound orbits around the  black hole with conformally coupled scalar hair. In section \ref{sec: precession orbit},  we will focus on the precessing orbits to give preliminary constraints  on charge parameters with the use of the orbits of S2 star in SgrA*.
We figure out samples of periodic orbit and discuss the effects charge parameters in section \ref{sec: periodic orbits}. The final section contributes to our conclusions and discussions. In this paper, we shall use the units $G = c = 1$. Additionally, in our theoretical evaluations, all the physical quantities are rescaled by the black mass parameter $M$ to be dimensionless, therefore, we set $M = 1$ unless we especially restate.

\section{Bound timelike orbits around  black hole with conformally coupled scalar hair}\label{sec:bound orbots}

The static and spherically symmetric metric for the spacetime with conformally coupled scalar hair derived from the action \eqref{eq:SZYL} is given by \cite{Astorino:2013sfa}
\begin{eqnarray}\label{eq:LM}
&&ds^2=-f(r)dt^2+\frac{1}{f(r)}dr^2+r^2(d\theta^2+\text{sin}^2\theta d\phi^2) \label{eq:LM},\\
&&\mathrm{with} ~~~ f(r)=1-\frac{2M}{r}+\frac{q^2}{r^2}+\frac{s}{r^2} \label{eq:fr}.
\end{eqnarray}
And the profiles of the matter fields are
\begin{eqnarray}\label{eq:chang}
\varphi=\pm\sqrt{\frac{6}{\kappa}}\sqrt{\frac{s}{q^2+s}}~~\text{and}~~A_\mu=-\frac{q}{r}\delta_\mu^t.
\end{eqnarray}
In the above solution, $M$,  $q$ and $s$ denote the mass parameter, electric charge parameter and conformally-coupled scalar charge parameter, respectively. According to these parameters' relation, the metric \eqref{eq:LM} and \eqref{eq:fr} can describe different spacetimes:
\begin{description}
  \item[(i)]  When $s>M^2-q^2$, $f(r)=0$ has no real root, so the metric describes a naked singularity without horizon.
  \item[(ii)] When $M^2-q^2\geq s\geq 0$, $f(r)=0$ has two positive real roots $r_{\pm}= M\pm\sqrt{M^2-q^2-s}$. So the metric describes a black hole with Cauchy horizon $r_-$ and event horizon $r_+$. {In particular, when $s=M^2-q^2$ the black hole becomes extreme with $r_{-}=r_{+}$, and the geometry is then a Reissner-Nordstr$\ddot{\text{o}}$m (RN) black hole with an electric charge $Q^2=q^2+s$.}
  \item[(iii)] When $-q^2<s<0$,  the profile of scalar field in \eqref{eq:chang} is imaginary,  {which is not considered in this article.}
  \item[(iv)] When $s<-q^2$, the metric also describes a black hole with two horizons $r_{\pm}$, in which the coefficient of $1/r^2$-term is negative differentiating from the RN black hole. Therefore, the black hole in this case is also dubbed a mutated RN black hole.
\end{description}
Thus, in this paper, we will work in the spherical black hole with comformally coupled scalar field of which the charge parameters should satisfy  $M^2-q^2\geq s\geq 0$ or $s<-q^2$. It is noted that the black hole spacetime \eqref{eq:LM}  is very close to the ``BBMB black hole'' constructed in \cite{Zou:2019ays, Bekenstein:1974sf}, {except that the scalar field here is regular instead of non-regular.} The Ricci scalar, the square of the Ricci tensor and the Kretschmann scalar for the black hole are then calculated as
\begin{align}
R &=g_{\mu \nu }R_{\mu \nu}=0,~~~R^2=R_{\mu \nu}R^{\mu \nu}=\frac{4 \left(q^2+s\right)^2}{r^8},\\
K^2 &=R_{\mu \nu \sigma \rho } R^{\mu \nu \sigma \rho }=\frac{8 \left(6 M^2 r^2-12 M q^2 r-12 M r s+7 q^4+14 q^2 s+7 s^2\right)}{r^8},
\end{align}
which denote that the charge parameters, $q$ and $s$, have influence on $R^2$ and $K^2$, but the Ricci scalar always vanishes.

Then, we shall start from the timelike geodesic around the spherically symmetric black hole \eqref{eq:LM}. To this end, we consider one massive particle moving on the equatorial plane ($\theta=\pi/2$) , and the corresponding Lagrangian is
\begin{equation}\label{eq:LLLL}
2 \mathcal{L}=g_{\mu \nu}\dot{x^{\mu}} \dot{x^{\nu}}=-f(r) \dot{t}^2+\frac{1}{f(r)}\dot{r}^2+r^2 \dot{\phi 
}^2 ,
\end{equation}
{with the dot represents the derivatives with respect to the affine parameter $\lambda$.}
Then the Euler-Lagrangian equation gives us
\begin{align}
p_{t} &=\frac{\partial \mathcal{L}}{\partial \dot{t}}=-f(r) \dot{t}=-E ,\label{eq:Pt}\\
p_{r} &=\frac{\partial \mathcal{L}}{\partial \dot{r}}=\frac{\dot{r}}{f(r)} ,\label{eq:Pr}\\
p_{\phi} &=\frac{\partial \mathcal{L}}{\partial \dot{\phi }}=r^2 \dot{\phi }=L ,\label{eq:Pphi}
\end{align}
where $E$ and $L$ denote the conserved energy and orbital angular momentum per unit mass of the particle. Then considering that $g_{\mu\nu}\dot{x^{\mu}}\dot{x^{\nu}}=-1$ for timelike geodesic,
we can derive the radial equation of motion as
\begin{equation}\label{jing}
\dot{r}^2=E^2-f(r)\left(1+\frac{L^2}{r^2}\right) ,
\end{equation}
from which we can get the effective potential of the radial motion of the particle
\begin{equation} \label{Veff}
V_{eff}\equiv E^2-\dot{r}^2=f(r)\left(1+\frac{L^2}{r^2}\right)=\left(1-\frac{2M}{r}+\frac{q^2}{r^2}+\frac{s}{r^2}\right)\left(1+\frac{L^2}{r^2}\right).
\end{equation}
It is obvious that the effective potential $V_{eff}$ depends explicitly on the angular momentum $L$, the radius $r$, the charge parameters $q$ and $s$.  In particular, $V_{\text{eff}}\mid_{r\to \infty}=1$ such that $E=1$ is the upper limit for the bound orbits because in this case  we have $\dot{r}^2=E^2-V_{\text{eff}}\mid_{r\to \infty}=0$.
The particle's motion is determined by the balance of the effective potential and the energy $E$. Therefore, according to \cite{2004graa.book}, we can then qualitatively discuss the properties of various possible motions of the particles around the black hole by analyzing the effective potential's profile, including the scattering orbits ($E>1$) for which particle can escape into infinity, and the bound orbits when $E\leq 1$.
Here we are interested in bound orbits for which the energy and angular momentum of the particle should be in certain region, as we will see later. Before analyzing the properties of general bound orbits, we will first study two particularly noteworthy bound orbits, i.e, the marginally bound orbits and innermost stable circular orbits  \cite{thorne2000gravitation}.

\begin{description}
  \item[$\bullet$ Marginally bound orbits (MBO) :]
 MBO  is characterized by a balance between the particle's kinetic energy and the gravitational potential energy, resulting in a zero radial velocity. Marginally bound orbits are of particular interest in black hole physics as they correspond to the innermost orbits a particle can have without falling into the black hole. It is a specific unstable circular orbit around the black hole, of which the maximum effective potential is $1$, so it yields to
\begin{equation} \label{Mbos}
V_{eff}=1, ~~\partial_{r} V_{eff}=0.
\end{equation}
Plugging Eq.\eqref{Veff} into Eq.\eqref{Mbos}, and adopting the convention $M=1$, we directly obtain two equations
\begin{align}
\frac{L^2 q^2}{r^4}+\frac{L^2 s}{r^4}-\frac{2 L^2}{r^3}+\frac{L^2}{r^2}+\frac{q^2}{r^2}+\frac{s}{r^2}-\frac{2}{r}=0, \label{mbolr} \\
-\frac{4 L^2 q^2}{r^5}-\frac{4 L^2 s}{r^5}+\frac{6 L^2}{r^4}-\frac{2 L^2}{r^3}-\frac{2 q^2}{r^3}-\frac{2 s}{r^3}+\frac{2}{r^2}=0. \label{mboLR}
\end{align}
Then, by simultaneously solving the above equations, we can derive the radial distance $r_{MBO}$ and the angular momentum $L_{MBO}$ of the particle around the black hole with conformally coupled scalar hair with respect to the charge parameters. The results are depicted in FIG.\ref{fig:mbos}. For positive scalar charge parameters, an increase in {$\lvert s\rvert$} leads to a gradual decrease in both the radius $r_
{MBO}$ and angular momentum $L_{MBO}$. Conversely, for negative scalar charge parameters, the trends in their variations are reversed. It is noteworthy that the blue, green, and red curves in the figure represent $q = 0$, $q = 1/4$, and $q = 1/3$, respectively, which show that for the black hole with larger $q$, the MBO always has smaller radius and angular momentum. It is noticed that the gap in the figure raises because the black hole exists only for $M^2-q^2\geq s\geq 0$ or $s<-q^2$ as we addressed in the previous section. 
%%%%%%%%%%%%
%%%%%%%%%%%%
\begin{figure} [h]
{\centering
\includegraphics[width=2.5in]{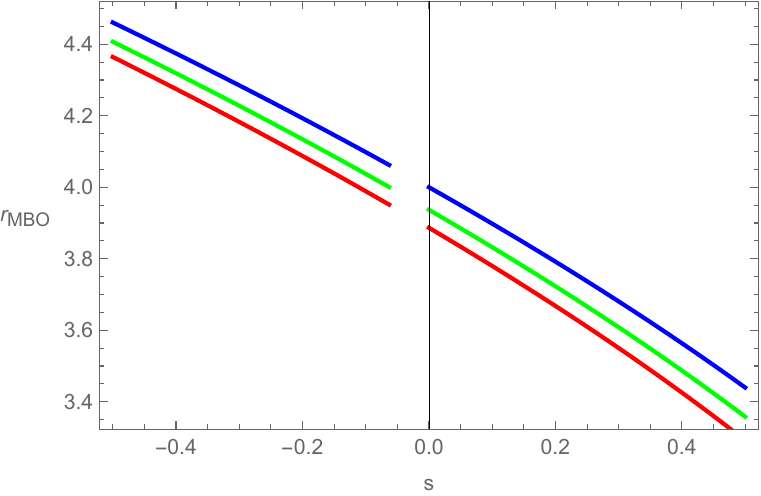}\hspace{0.5cm}
\includegraphics[width=3in]{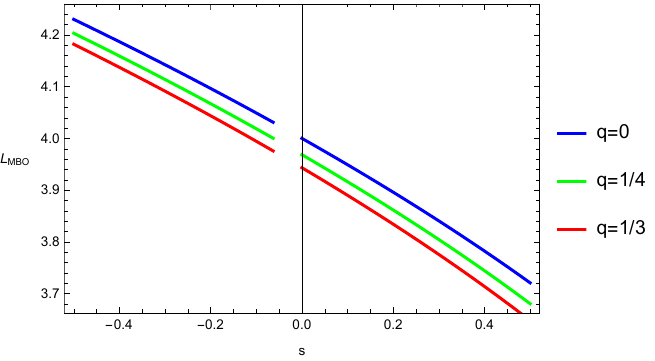}
\caption{ The radius and  angular momentum  as functions of scalar charge parameter for the particles in the marginally bound orbit around the black hole with conformally coupled scalar hair. Here we choose samples of electric charge parameters. }
   \label{fig:mbos}}
\end{figure}

\item[$\bullet$ Innermost stable circular orbit (ISCO):] 
ISCO is  the closest stable circular orbits that a particle can orbit the black hole. So a small perturbation could not cause the particle to either fall inward or escape outward. ISCO encodes key information of the dynamics of matter and energy in extreme gravitational fields, so it  is of great importance in black hole physics such that studying accretion disks around black holes, as well as for modeling and predicting gravitational wave signatures from merging black holes. ISCO should satisfy the following conditions,
\begin{equation} \label{ISCOs}
V_{eff}=E^2,~~\partial_{r} V_{eff}=0,~~\partial^{2}_{r}V_{eff}=0,
\end{equation}
which are deduced as 
\begin{align}
\frac{L^2 q^2}{r^4}+\frac{L^2 s}{r^4}-\frac{2 L^2}{r^3}+\frac{L^2}{r^2}+\frac{q^2}{r^2}+\frac{s}{r^2}-\frac{2}{r}-E^2+1=0,\\
-\frac{2 L^2 q^2}{r^5}-\frac{2 L^2 s}{r^5}+\frac{3 L^2}{r^4}-\frac{L^2}{r^3}-\frac{q^2}{r^3}-\frac{s}{r^3}+\frac{1}{r^2}=0,\\
\frac{10 L^2 q^2}{r^6}+\frac{10 L^2 s}{r^6}-\frac{12 L^2}{r^5}+\frac{3L^2}{r^4}+\frac{3 q^2}{r^4}+\frac{3 s}{r^4}-\frac{2}{r^3}=0.
\end{align}
By solving the above equation group, we can extract the key information of the ISCO. The radius $r_{{ISCO}}$, angular momentum $L_{ISCO}$ and energy $E_{ISCO}$ with respect to the scalar charge $s$ for different electric charges $q$ are plotted in FIG.\ref{fig:isco}, which shows that $r_{{ISCO}}$, $L_{ISCO}$ and $E_{ISCO}$ all decrease as the increasing of both charge parameters in the allowed parameter space.
%%%%%%%%%%%%
%%%%%%%%%%%%
\begin{figure} [h]
{\centering
\includegraphics[width=2.15in]{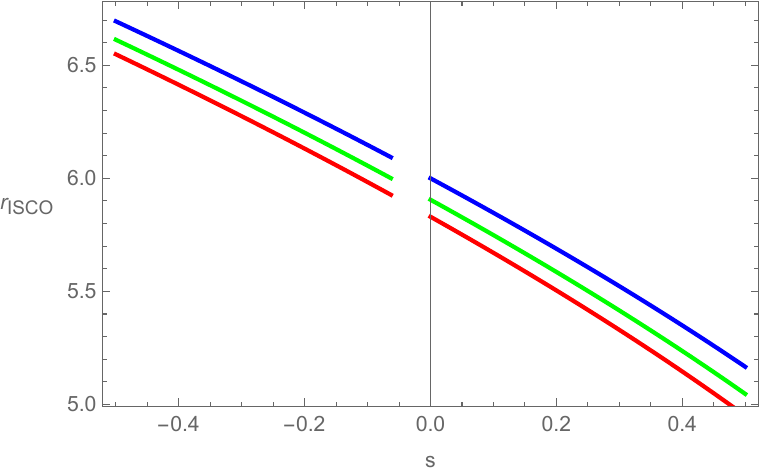}\hspace{0.1cm}
\includegraphics[width=2.15in]{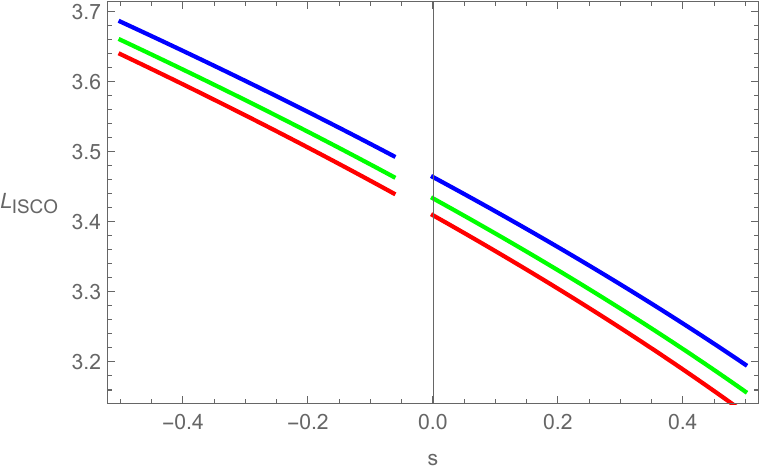}\hspace{0.1cm}
\includegraphics[width=2.55in]{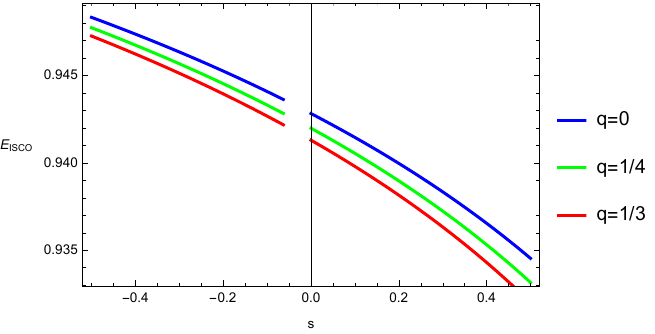}
\caption{ The radius, angular momentum  and energy as functions of scalar charge parameter for the particles in the ISCO around the black hole with conformally coupled scalar hair.}
\label{fig:isco}}
\end{figure}
  %%%%%%%%%%%%
%%%%%%%%%%%%
\end{description}

\begin{figure} [h]
{\centering
\includegraphics[width=5in]{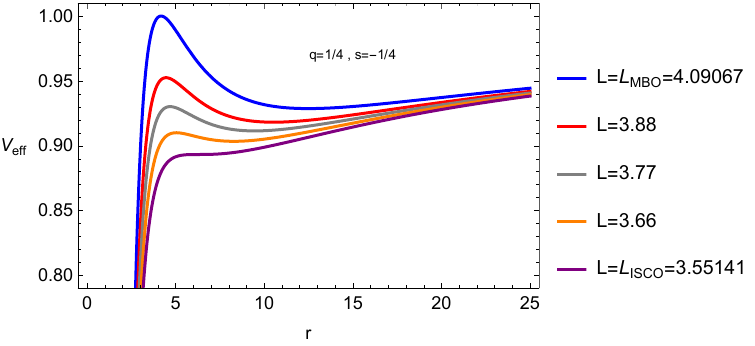} \hspace{0.5cm}
   \caption{ The effective potential of the particles in the bound orbits. The blue curve is for the
MBO with $L = 4.09067$ which has two extremal points, while the purple curve is for the ISCO with $L = 3.55141$ which has one extremal point. Here we fix the charge parameters as $q=1/4$ and $s=-1/4$.}   \label{fig:Veff}}
\end{figure}

\begin{figure} [h]
{\centering
\includegraphics[width=3in]{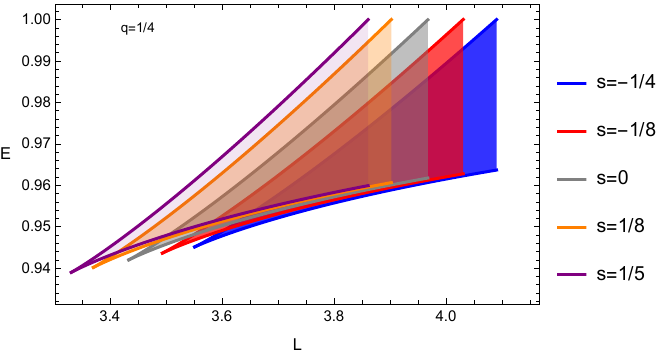}\hspace{0.5cm}
\includegraphics[width=3in]{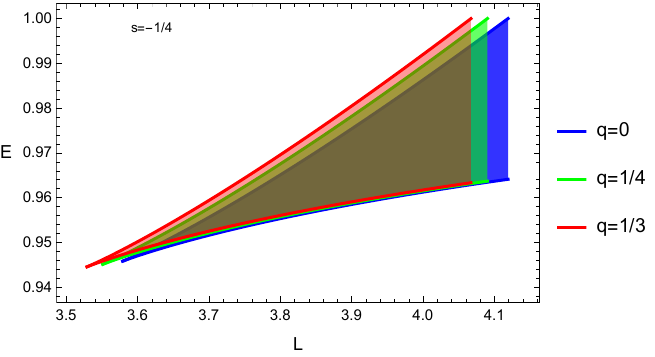}
   \caption{ The allowed $(L,E)$ regions for the bound timelike orbits around the black holes with selected charge parameters.}   \label{fig:EL}}
\end{figure}

\begin{figure} [h]
{\centering
\includegraphics[width=5in]{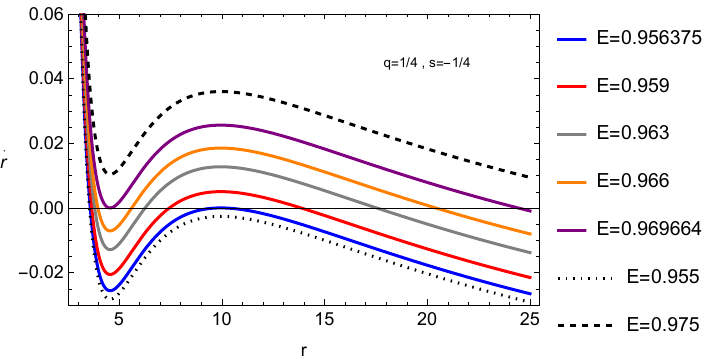} \hspace{0.5cm}
   \caption{ The general radial motion $\dot{r}^2$ as a function of $r$ around the hairy black hole with $q = 1/4$ and $s=-1/4$. Here we choose $L=\frac{L_{BMO}+L_{ISCO}}{2} = 3.82104$ and tune the energies of the particle. The bound orbits only exist when $\dot{r}^2= 0$ has at least two roots. More details can be seen in the main text.}   \label{fig:RD}}
\end{figure}

Then, we shall analyze the properties of general bound orbits. Since the bound orbit only exists when the particle's energy is no more than 1 as we previously mentioned, so here we shall focus on the cases  between MBO and ISCO to discuss the properties of bound orbits. The analysis mainly involves  studying the radial motion of timelike particles and profiles of its effective potential. For the black hole with fixed parameters, the angular momenta for the general bound orbits should be between the corresponding $L_{ISCO}$ and $L_{MBO}$, and the particle in a bound orbit with selected angular momentum can only have certain region of energy.  To explicitly show these properties, we show the sample of effective potential for various bound orbits with $q=1/4$ and $s=-1/4$ in FIG. \ref{fig:Veff}, which has two extremal points moving closer to each other as the angular momentum decreases from $L_{MBO} = 4.09067$ and finally merging together when it reaches  $L_{ISCO}$. The orbits with angular momentum between $L_{ISCO}$ and $L_{MBO}$ could be bound by the corresponding potential well, depending on the particle's energy. 
As shown in FIG. \ref{fig:EL}, we figure out the allowed $(L-E)$ space for the bound orbits. It shows that the two charge parameters will  both {alter} the allowed $(L-E)$ region of bound orbits.
Then, we shall analyze the radial motions in the bound orbits for the particles. The sample results with fixed {$L=(L_{BMO}+L_{ISCO})/2$} for different energies are shown in FIG. \ref{fig:RD}. {In this case, only the particles with the energy satisfying $0.956375  \leq E \leq 0.969664$ can have the bound orbits, where $\dot{r}^{2}=0$ has at least two roots.} In details, for the purple curve (the first solid line from top) with $E=0.969664$, {$\dot{r}^{2}=0$ has only two roots and} the bound orbits exist in the regime between the two roots. Then for the particle with smaller energy, $\dot{r}^2=0$ has three roots and  the bound orbits exist in the regime { with $\dot{r}^2>0$} between the last two roots, and {the bound orbits converge} to the ISCO with $E=0.956375$. Additionally, when the particle's energy is beyond the aforementioned regime,  $\dot{r}^2=0$ has no or single root {so no bound orbits exist}.

Next, we shall study the special subclasses of bound orbits, the precessing and periodic orbits, around the hairy black hole with conformally coupled scalar field.

\section{Orbital pericenter precession and constraints on charge parameters from S2 star's motion}\label{sec: precession orbit}
Due to the spherically symmetric nature of the background black hole, the motion of particles is solely determined by the radial coordinate $r$ and the azimuthal angle $\phi$.
For a precessing orbit with two turning points $r_1$ and $r_2$, the timelike particle {will be} reflected between the turning points. So, the apsidal angle $\Delta \phi$ passed by the particle in this process is
\begin{eqnarray}\label{eq:Dleta-phi}
\Delta \phi&=&\oint d\phi=2\int_{r_1}^{r_2}\frac{d\phi}{dr}dr.
\end{eqnarray}
Especially, for the periastron precession, the particle's trajectory is described by   \cite{2004graa.book}
\begin{eqnarray}\label{eq:r(Psi)}
r=\frac{a(1-e^2)}{1+e\cos \Psi} ,
\end{eqnarray}
of which the turning points are the periastron, $r_p=a(1-e)$, and apastron, $r_a=a(1+e)$, respectively. Here $a$ and $e$ are the semi-major axis and the eccentricity
of the orbit, while $\Psi$ denotes the intersection angle between the semi-major axis and radial of the orbit. Subsequently,  the above angle $\Delta\phi$ in this  case is evaluated as
\begin{eqnarray}\label{eq:Delta-phi}
\Delta \phi=2\int_{r_p}^{r_a}\frac{d\phi}{dr}dr=2\int_{0}^{\pi}\frac{d\phi}{d\Psi}d\Psi ,
\end{eqnarray}
with
\begin{eqnarray}\label{eq:dphidPsi}
\frac{d\phi}{d\Psi}=\frac{d\phi/dr}{d\Psi/dr}=
\frac{a e(1-e^2)L\sin\Psi}{r^2(1+e\cos\Psi)^2\sqrt{E^2-f(r)\left(1+\frac{L^2}{r^2}\right)}},
\end{eqnarray}
where the expression of $d\phi/dr$ {is} derived from \eqref{eq:Pphi} and \eqref{jing}, {and} $d\Psi/dr$ {is} obtained from \eqref{eq:r(Psi)}.
Recalling  that the turning points {satisfy} $\dot{r}|_{r=r_p}=\dot{r}|_{r=r_a}=0$,  we can obtain
 \begin{equation} \label{EELL}
 E^2=\frac{(r_p^2-r_a^2) f(r_p)f(r_a)}{r_p^2f(r_a)-r_a^2f(r_p)},~~~L^2=\frac{r_p^2r_a^2( f(r_p)-f(r_a))}{r_p^2f(r_a)-r_a^2f(r_p)}.
 \end{equation}

Then, to perform the integral in \eqref{eq:Delta-phi}, we shall do some approximation on the integral function \eqref{eq:dphidPsi}. We consider the weak gravitational field case, i.e., $M/r$ is treated as small quantity. Then we rescale the charge parameters as $Q^2=\frac{q^2}{M^2}$ and $S=\frac{s}{M^2}$, {which are also considered as small quantities}.
Subsequently, we can obtain
\begin{equation} \label{dphidpsi}
\begin{aligned}
\frac{d\phi}{d\Psi} = & 1 + \frac{(3 + e \cos \psi)M}{a(1 - e^2)} \\
& +  S \left(\frac{M}{2a^2(e^2-1)}-\frac{M^2(12+7e \cos\psi+e^2 \cos^2 \psi)}{2 a^2 (e^2-1)^2}\right)
 +  Q^2 \left(\frac{M}{2a^2(e^2-1)}-\frac{M^2(12+7e \cos\psi+e^2 \cos^2 \psi)}{2 a^2 (e^2-1)^2}\right) +\cdots,
\end{aligned}
\end{equation}
in which the first line  represents the result for Schwarzschild black hole in GR.
Then, substituting \eqref{dphidpsi} into \eqref{eq:Delta-phi}, we will obtain the precessing angle as
\begin{align}
\triangle \omega _{hairy} & =\triangle \phi-2\pi=\frac{6 \pi  M}{a(1-e^2)}+S \left(\frac{2a(e^2-1)\pi M-(24+e^2)\pi M^2}{2a^2 (-1+e^2)^2}\right) +Q^2 \left(\frac{2a(e^2-1)\pi M-(24+e^2)\pi M^2}{2a^2 (-1+e^2)^2}\right),
  \end{align}
which will recover the result for Schwarzschild black hole, $\triangle \omega _{GR}=\frac{6 \pi  M}{a-a e^2}$, {if} the charge parameters vanish.

Now, we are ready to use the observations of Schwarzschild precession of the S2 star around Sgr A* to test the current theory.
The Schwarzschild precession of the S2 star around Sgr A* was measured by the GRAVITY project with the technics of  spectroscopic
and astrometric measurements{, which is the first measurement of periastron  precession for a star around supermassive black hole} \cite{GRAVITY:2020gka}. The best-fit ratio between the measured result to that predicted by GR is given by
\begin{eqnarray}\label{eq:fsp}
f_{sp}\equiv \frac{\Delta \omega_{s2}}{\Delta \omega_{GR}}=1.1\pm 0.19,
\end{eqnarray}
within $1\sigma$ uncertainty which leaves some space for alternative gravity beyond GR.  So, we shall presuppose that the central supermassive black hole in SgrA* is the hairy black hole with conformally coupled scalar field, and the precessing motion of s2 star follows the trajectory we studied above. Subsequently, the shift $f_{sp}$ is then evaluated by
\begin{equation} \label{fsp1}
 f_{sp}=\frac{\triangle \omega _{hairy}}{\triangle \omega _{GR}}=1-S \left(\frac{1}{6}-\frac{(24+e^2) M}{12a(e^2-1)}\right) -Q^2 \left(\frac{1}{6}-\frac{(24+e^2) M}{12a(e^2-1)}\right).
\end{equation}

\begin{figure} [h]
{\centering
\includegraphics[width=3in]{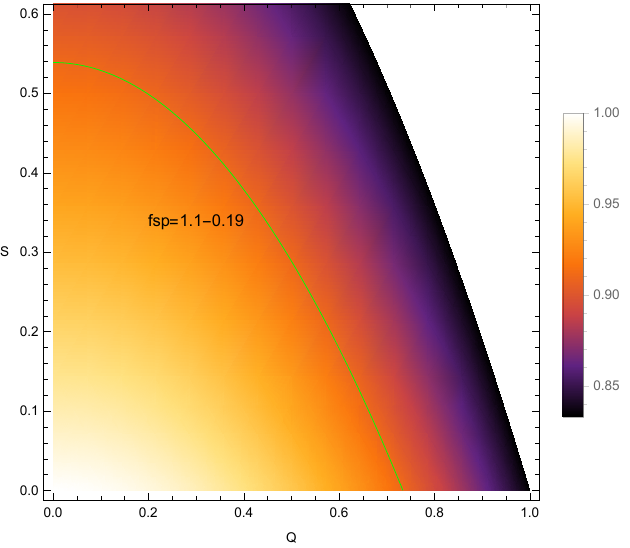}\hspace{0.5cm}
\includegraphics[width=3in]{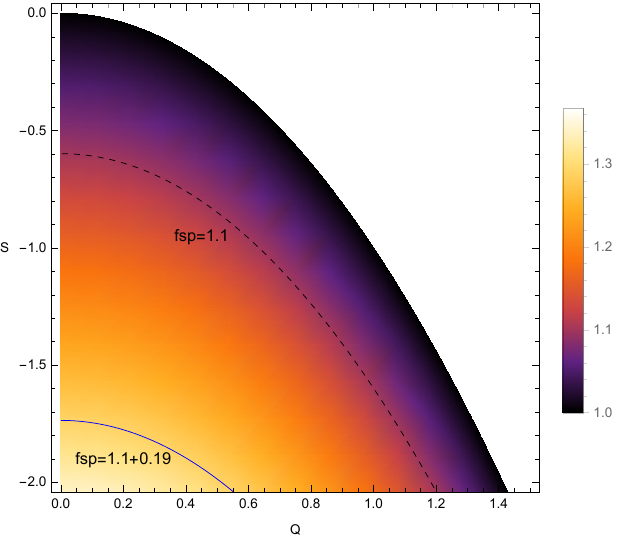}
   \caption{ The density plot of $f_{sp}$ of the S2 star's motion in the parameter $(q/M,s/M^2)$ space by treating the current hairy black hole as the central supermassive black hole in SgrA*.  The left plot shows the case with $S> 0$, while the right one illustrates the case with $S < 0$.}   \label{fig:sq}}
\end{figure}

Then we can calculate $f_{sp}$ by considering the realistic values of various physical quantities in the S2 star's motion, such as the mass of SgrA*, $M_{SgrA}=4.3\times10^6 M_{\odot}$, the luminosity distance  between the S2 star and SgrA*, $D=8.35\, kpc$, and the eccentricity and the semi-major axis of the orbit $e=0.884649$ and $a=125.058\,mas$ \cite{GRAVITY:2020gka}.  The results in $(S,Q)$ parameter space are  depicted in FIG.\ref{fig:sq}, the left of which we focus on hairy black hole with $M^2-q^2\geq s\geq 0$ while the right panel corresponds to $s<-q^2$. It is obvious that the S2 star's motion can further constrain the charge parameters. In details, for $s\geq 0$ in the left plot, only the charge parameters in the lower left region of the green solid curve, which corresponds to lower bound $f_{sp}=1.1-0.19$, match the S2 star's observation, while for $s<-q^2$ in the right plot, the charge parameters are constrained  into the upper right region of the blue solid curve which gives the upper bound $f_{sp}=1.1+0.19$.  By carefully comparing the constrained parameter region from black hole shadow addressed in \cite{Khodadi:2020jij}, we find that our constraints on the charge parameters from S2 star's motion are tighter.

\section{Effects of the charge parameters on periodic motions}\label{sec: periodic orbits}
\begin{figure} [h]
{\centering
\includegraphics[width=2.5in]{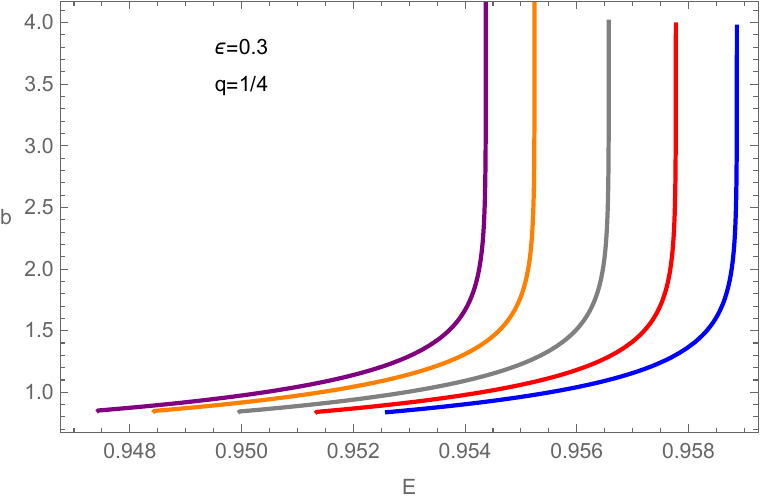}\hspace{0.5cm}
\includegraphics[width=3.1in]{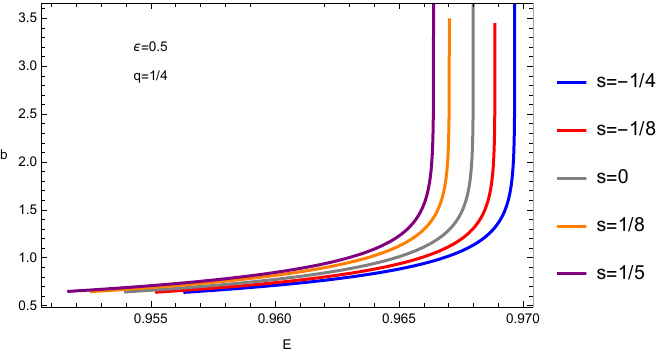}
\includegraphics[width=2.5in]{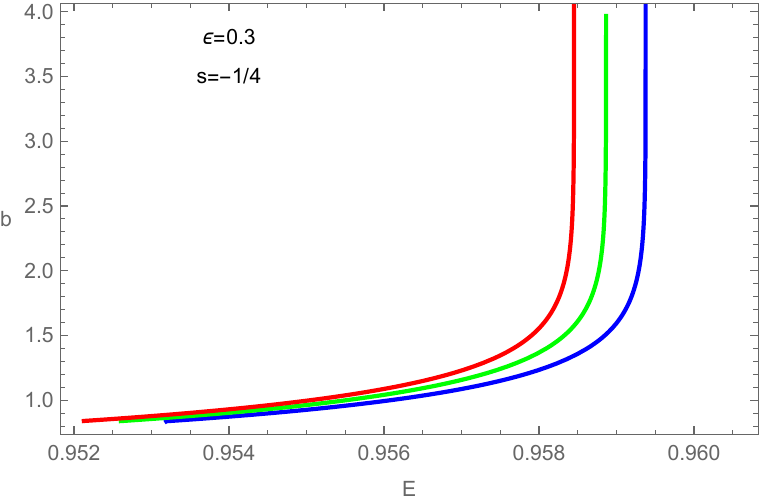}\hspace{0.5cm}
\includegraphics[width=3in]{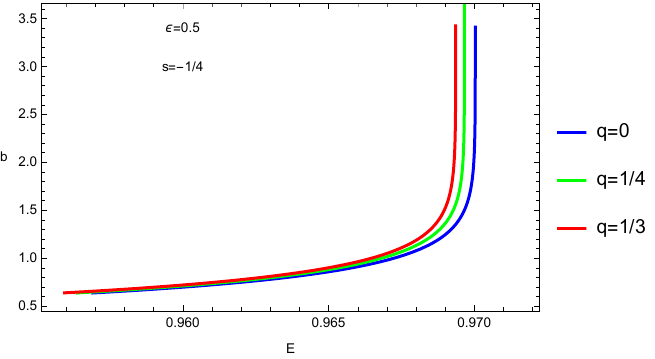}
   \caption{The number $b$ of periodic bound orbit as a function of energy $E$ for selected black hole parameters. We choose two angular momenta by setting $\epsilon=0.3$ (left column) and $\epsilon=0.5$ (right column).}   \label{fig:be(q)}}
\end{figure}

\begin{figure} [h]
{\centering
\includegraphics[width=2.5in]{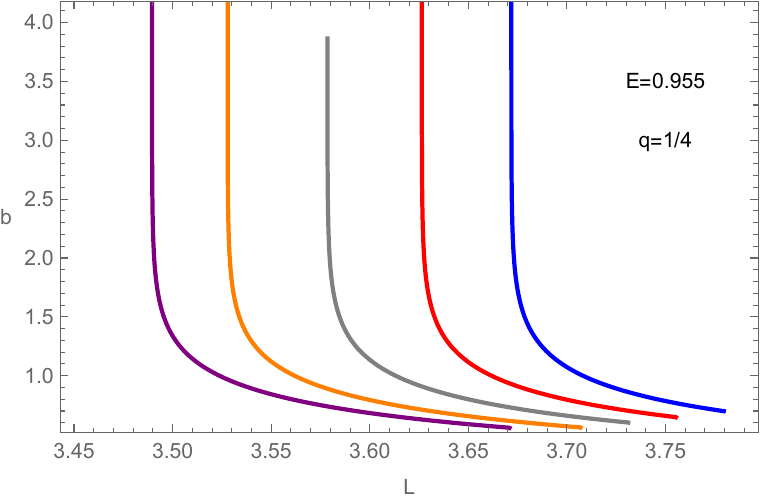}\hspace{0.5cm}
\includegraphics[width=3in]{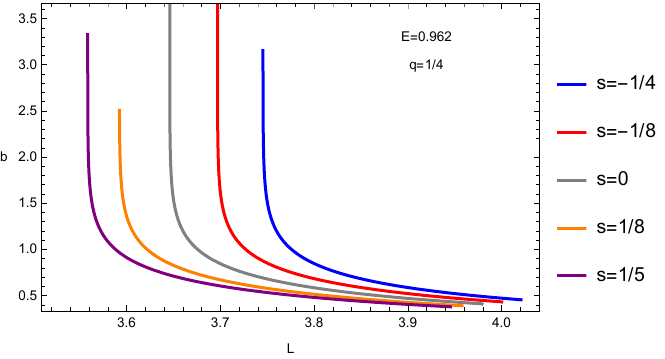}
\includegraphics[width=2.5in]{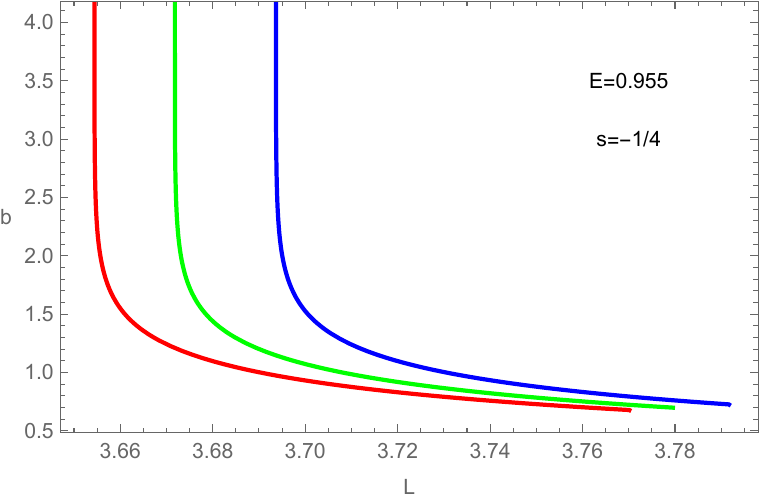}\hspace{0.5cm}
\includegraphics[width=3in]{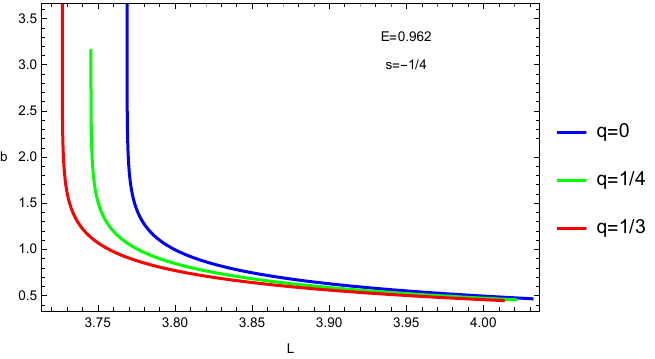}
   \caption{The number $b$ of periodic bound orbit as a function of the angular momentum $L$ for selected black hole parameters.  We consider two values of energy, $E=0.955$ (left column) and $E=0.962$ (right column).}   \label{fig:be(s)}}
\end{figure}

In this section, we will move on to study the periodic orbits, which repeat themselves over time. They describe the regular and predictable motion of objects in space, ranging from planets orbiting the Sun to stars moving within galaxies. The study of periodic orbits is crucial for understanding the stability and dynamics of celestial systems, as well as for making predictions about their long-term behavior. As suggested in \cite{Levin:2008mq}, the bound periodic orbit can be described by a unique rational number $b$ as
\begin{equation}
b=\frac{\triangle \phi }{2 \pi }-1=w+\frac{v}{z} ,
\end{equation}
where the angle $\Delta \phi$ is defined in \eqref{eq:Dleta-phi}, and the integers $z$, $w$ and $v$ are the zoom number, whirl number and  vertex number in the orbit, respectively. Therefore, the values of $b$ can be further calculated via
\begin{equation} \label{eq:b1}
b=\frac{\triangle \phi }{2 \pi }-1=\frac{1}{\pi} \int_{r_1}^{r_2} \frac{d\phi}{dr} dr=\frac{1}{\pi} \int_{r_1}^{r_2} \frac{L}{r^2 \sqrt{E^2-f(r)(1+\frac{L^2}{r^2})}} dr-1.
\end{equation}
We show the results of $b$ for periodic orbits as functions of energy $E$ as well as angular momentum $L$ for selected charge parameters in FIG.\ref{fig:be(q)} and FIG.\ref{fig:be(s)}. It is noticeable that here for better comparison, we have introduced a new quantity  {$\epsilon\equiv(L-L_{ISCO})/(L_{MBO}-L_{ISCO})\in [0,1]$} because the angular momentum for bound orbit should satisfy $L_{ISCO}<L<L_{MBO}$.
In FIG.\ref{fig:be(q)}, we choose $\epsilon=0.3$ and $\epsilon=0.5$ to describe $b$ as a function of energy. For all curves,  $b$ first increases slowly with the increase of energy $E$ and then experiences a sudden surge when $E$ reaches a maximal value, but larger $\epsilon$  suppress the value of $b$. Moreover, we find that increasing both charge parameters will shift the curve into smaller energy.
In FIG. \ref{fig:be(s)}, we fix $E=0.955$ and $E=0.962$, respectively, to  describe  $b$ as function of angular momentum. For each curve with chosen charge parameters, as the angular momentum decreases, $b$
increases smoothly and then explodes as $L$ approaches its minimal values. Similarly, for black hole with both larger charge parameters, the curve is shifted into left corresponding smaller angular momentum of the orbits.

\begin{figure} [h]
{\centering
\includegraphics[width=7in]{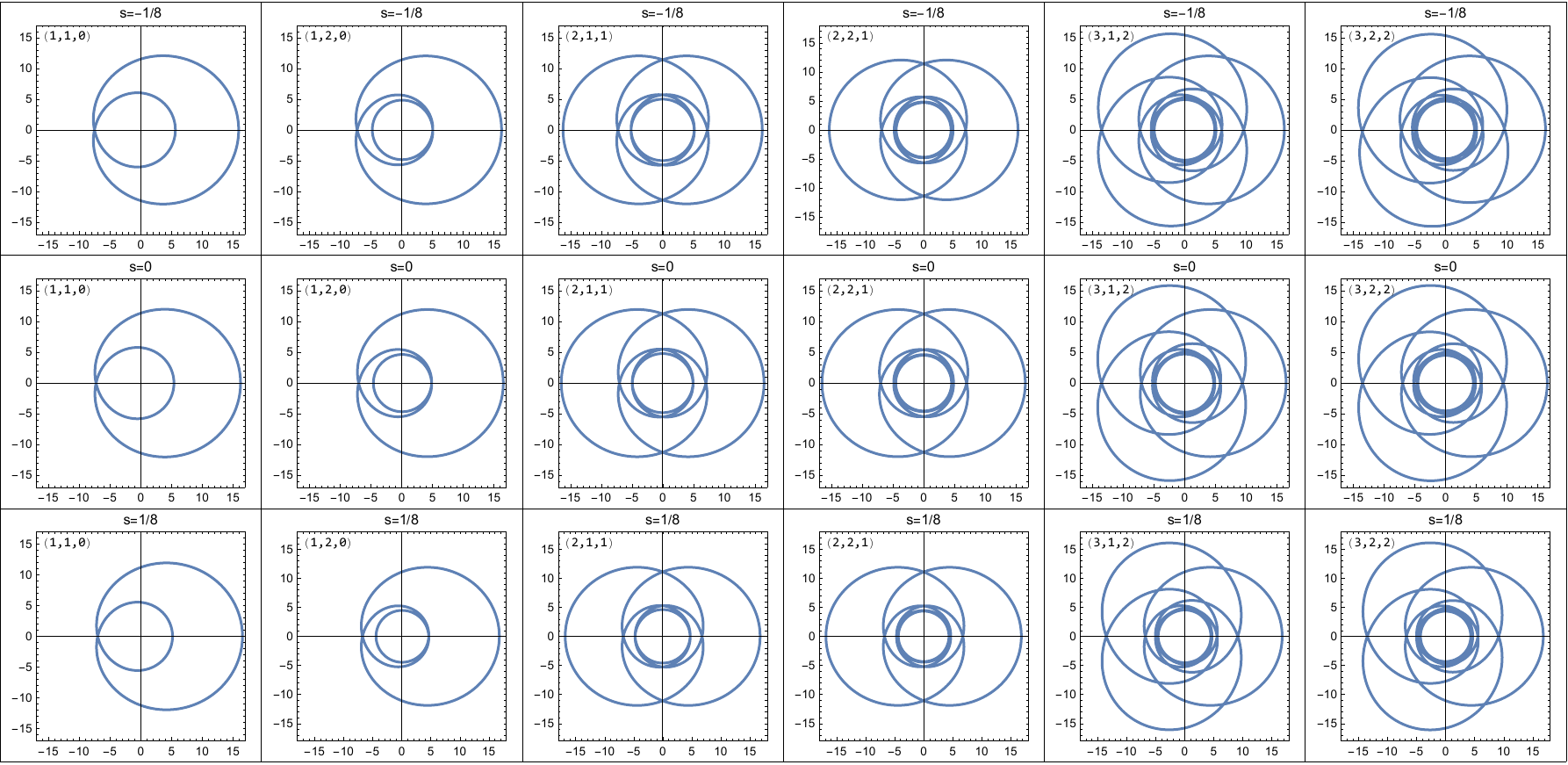}
   \caption{The trajectories of the periodic orbits ($z$,$w$,$v$)  with selected scalar charges by  the electric charge $q=1/4$. We fix $E=0.96$ for all orbits and the corresponding $L$ were listed in Table.\ref{table02}. }  \label{fig:s-o3}}
\end{figure}

\begin{figure} [h]
{\centering
\includegraphics[width=7in]{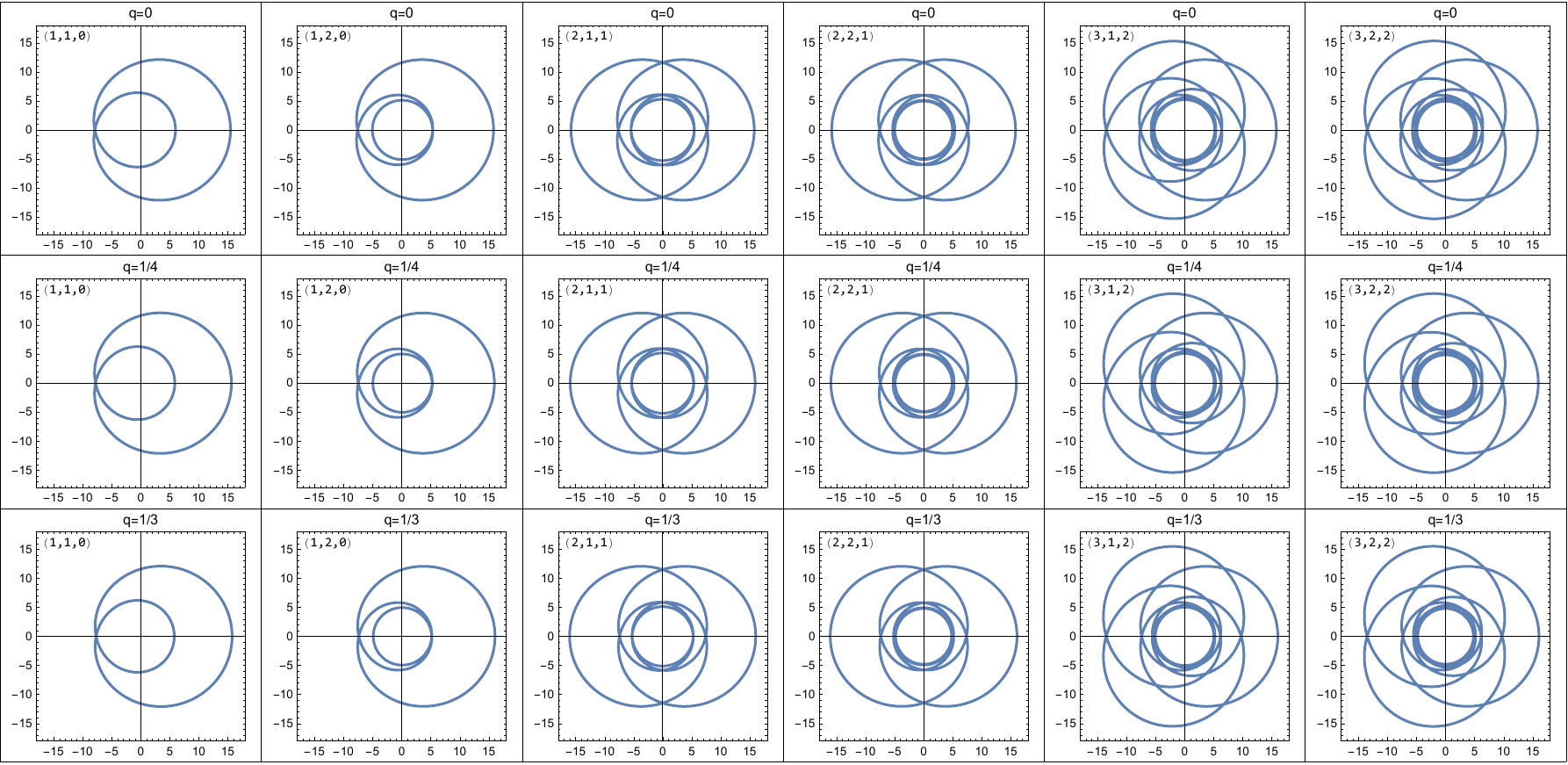}
   \caption{The trajectories of the periodic orbits ($z$,$w$,$v$)  with selected electric charges by  the scalar charge $s=-1/4$. We fix $E=0.96$ for all orbits and the corresponding $L$ were listed in Table.\ref{table02}. }  \label{fig:q-o}}
\end{figure}

Then in order to figure out the trajectories of the periodic orbit $(z,w,v)$ around the black hole with conformally coupled scalar field, we shall first determine the corresponding energy $E_{(z,w,v)}$ and angular momentum $L_{(z,w,v)}$. To this end, we firstly list the corresponding energy $E_{(z,w,v)}$ with fixed $L=\frac{L_{MBO}+L_{ISCO}}{2}$ (i.e., $\epsilon=0.5$) in Table \ref{table01}, and the corresponding angular momentum $L_{(z,w,v)}$ with fixed $E=0.96$ in Table \ref{table02}, respectively. The effects of charge parameters indeed follow the findings in the previous figures of $b$.
In details, as the electric charge $q$ increases, both $E_{(z,w,v)}$ and $L_{(z,w,v)}$ gradually decrease. While when $s > 0$, both $E_{(z,w,v)}$ and $L_{(z,w,v)}$ gradually increase with the increasing amount of scalar hair, but when $s < 0$, the situation is exactly the opposite. This suggests that, under otherwise identical conditions, a massive particle orbiting around a hairy black hole in current model will have higher energy $E$ and angular momentum $L$ due to the presence of the electric charge $q$ compared to that in Schwarzschild case. While the effect of the scalar hair $s$ on the energy $E$ and angular momentum $L$ of the massive particle compared to the Schwarzschild black hole depends on its sign.
With the data in hands, we can figure out the periodic orbits with $(z,w,v)$ in the polar coordinates $(r,\phi)$. By selecting several charge parameters, we show the exemplified results with fixed $E=0.96$ in FIG. \ref{fig:s-o3} and FIG. \ref{fig:q-o} of which the corresponding $L$ is present in Table \ref{table02}. It is obvious that the zoom number $z$ describes the number of the leaf pattern of the orbit,  the whirl number $w$ describes  the number of nearly circular whirls close to periastron per leaf, and the vertex number $v$ gives the order of which the leaves are traced out.  It is found by comparing the orbits with different $(z,w,v)$ in each row that for orbits with larger $z$, the leaf pattern grows such that the orbit becomes more complicate. Moreover, by comparing the orbits with the same $(z,w,v)$  in each column, we see that both larger charge parameters will correspond to  larger outermost trajectory in the orbit.

\begin{table}[]
\begin{tabular}{|c|c|c|c|c|c|c|c|}
\hline
\null
& & $E_{(1, 1, 0)}$ & $E_{(1, 2, 0)}$ & $E_{(2,1,1)}$ & $E_{(2,2,1)}$ & $E_{(3,1,2)}$ & $E_{(3,2,2)}$ \\ \hline
\multirow{5}{*}{q=1/4}&
s=-1/4 & 0.96683639 & 0.96961000 & 0.96928037 & 0.96965657 & 0.96946435 & 0.96966032 \\ \cline{2-8}
&s=-1/8 & 0.96592008 & 0.96881088 & 0.96846442 & 0.96886070 & 0.96865754 & 0.96886472 \\ \cline{2-8}
&s=0 & 0.96489999 & 0.96793317 & 0.96756604 & 0.96798662 & 0.96777019 & 0.96799097 \\ \cline{2-8}
&s=1/8 & 0.96376189 & 0.96696084 & 0.96656826 & 0.96701885 & 0.96678592 & 0.96702364 \\ \cline{2-8}
&s=1/5 & 0.96300537 & 0.96632358 & 0.96897195 & 0.96638494 & 0.96614004 & 0.96639003 \\ \hline \hline
\multirow{3}{*}{s=-1/4}&q=0 & 0.96726202 & 0.96998344 & 0.96966159 & 0.97002902 & 0.96984153 & 0.97003267 \\ \cline{2-8}
&q=1/4 & 0.96683635 & 0.96960971 & 0.96928036 & 0.96965657 & 0.96946433 & 0.96966033 \\ \cline{2-8}
&q=1/3 & 0.96649082 & 0.96930758 & 0.96897195 & 0.96935553 & 0.96915928 & 0.96935938 \\ \hline
\end{tabular}
\caption{The energy $E$ for the samples of periodic orbits ($z$,$w$,$v$) with selected charge parameters. Here we fix the particle's energy as $L=\frac{L_{MBO}+L_{ISCO}}{2}$ or $\epsilon=0.5$.
\label{table01} }
\end{table}

\begin{table}[]
\begin{tabular}{|c|c|c|c|c|c|c|c|}
\hline
& & $L_{(1, 1, 0)}$ & $L_{(1, 2, 0)}$ & $L_{(2,1,1)}$ & $L_{(2,2,1)}$ & $L_{(3,1,2)}$ & $L_{(3,2,2)}$ \\ \hline
\multirow{5}{*}{q=1/4}
&s=-1/4 & 3.75673553 & 3.72589008 & 3.73023371 & 3.72514433 & 3.72789901 & 3.72507507 \\ \cline{2-8}
&s=-1/8 & 3.70853363 & 3.67815973 & 3.68239449 & 3.67744283 & 3.68011203 & 3.67737619 \\ \cline{2-8}
&s=0 & 3.65802749 & 3.62800556 & 3.63215779 & 3.62731074 & 3.62991465 & 3.6272476 \\ \cline{2-8}
&s=1/8 & 3.60488873 & 3.57507621 & 3.57917598 & 3.57439597 & 3.57695756 & 3.57433458 \\ \cline{2-8}
&s=1/5 & 3.57157418 & 3.54180416 & 3.54588981 & 3.541128547 & 3.54367764 & 3.54106775 \\ \hline
\hline
\multirow{3}{*}{s=-1/4}
&q=0 & 3.78005632 & 3.74893605 & 3.75334300 & 3.74817344 & 3.75097801 & 3.74810214 \\ \cline{2-8}
&q=1/4 & 3.75673575 & 3.72589010 & 3.73023382 & 3.72514433 & 3.72789899 & 3.72507507 \\ \cline{2-8}
&q=1/3 & 3.73824671 & 3.70759710 & 3.71189563 & 3.70686335 & 3.70958252 & 3.70679552 \\ \hline
\end{tabular}
\caption{The angular momentum  $L$ for the samples of periodic orbits ($z$,$w$,$v$) with selected charge parameters. Here We fix the energy of the particle as $E=0.96$.
\label{table02} }
\end{table}

\newpage

\section{Closing remarks}

Searching for alternative theories beyond GR still attracts the attention of physicists, especially by introducing additional scalar fields into the action of GR. An intuitive approach for investigating partial properties of such gravity theories is to study the appearance of the black hole images in specified spacetime backgrounds. In this article, we have considered a static spherically symmetric black hole with the conformally coupled scalar field, and study the properties of the bound timelike geodesics, precessing and the periodic orbits. Our results suggest that the scalar and electric charges have significant influences on the motions of the timelike particles. We also provide tighter constraints on the charge parameters from the observational result of the S2 star's precession in SgrA* than those from the black hole shadow.

We firstly investigated the general properties of bound orbits of the timelike particles, which exist between the MBOs and the ISCOs, in the considered hairy black hole. We demonstrate that under the specified parameter condition, the increasing of scalar and charge parameters will both gradually decrease the radius of MBOs and ISCOs and angular momentum of the particles on these orbits. The allowed $(L-E)$ regions for the bound orbits are then altered by the increasing parameters.

Then we studied the orbital pericenter precession and constraint on charge parameters by employing the observational results from S2 star's motion. The error margin of the ratio between the measured precessing angle to that predicted by GR leaves some space for alternative gravity beyond GR. By considering the realistic values of S2 star's motion, we have found that in the weak field case the constraints on the charge parameters are tighter than the constraints from the black hole shadow. 

Finally, we investigated the effects of the scalar and charge parameters on the periodic orbits. The results indicate that under the identical conditions a massive particle orbiting around the considered hairy black hole will have smaller energy and angular momentum with a larger charge parameter. These conclusions are also consistent with the previous sections. We also illustrated the trajectories of the periodic orbits and the larger parameters correspond to larger outermost trajectory  in the orbits. In general, during the  periodic motions, the particle's energy and angular momentum will decrease, which is accompanied by the emission of gravitational waves. Thus, the periodic orbits play important role in the study of gravitational wave emitted in the process of the two initial black holes of extreme mass ratio approaching to each other (EMRI), which was firstly proposed in \cite{Glampedakis:2002prd} and further emphasized in \cite{Levin:2008mq}. More recently, by considering the central black hole as a polymer black hole in loop quantum gravity \cite{Tu:2023xab} and Einsteinian cubic black hole \cite{Li:2024tld}, the connections  between the periodic orbits of timelike particles and the gravitational wave radiated from EMRI were further preliminarily studied. So along this line, based on the current work, it is interesting to study the effects of charge parameters on the gravitational radiation in the EMRI system via the  periodic orbits around the hairy black hole. Moreover, it is also of interest to explore the possible effect of  this gravitational radiation on the evolution of periodic orbits \cite{Mino:1996nk}.

Using black hole images to study the characteristics of different gravitational theories has always been one of the most concise and intuitive methods to distinguish between GR and alternative theories. With the development of observation technology, more and more details of black hole images will bring us increasingly strict parameter constraints. This will provide evidence support for further excluding certain theories of gravity. Although in this work we have investigated the timelike bound orbits and pericenter precession around black hole with conformally coupled scalar hair, there are still other directions that could provide more insights about the current gravitational theory, for example, black hole  illuminated by thin accretion disks or immersed in dark matter halo, which we will leave to future work.

\begin{acknowledgments}
We appreciate Yuan Meng and Xi-Jing Wang for helpful discussions. This work is partly supported by Natural Science Foundation of China under Grants Nos.12375054, 12005184, 12175192, and Natural Science 
Foundation of Jiangsu Province under Grant No. BK20211601.
\end{acknowledgments}

\bibliography{ref}
\bibliographystyle{apsrev}

\end{document}